\documentclass[onecolumn,amsmath,amssymb]{qm2008_abs}
\usepackage{graphicx}
\usepackage{dcolumn}
\usepackage{bm}

\usepackage{epsfig}

\def\bea{\begin{eqnarray}}  \def\eea{\end{eqnarray}}
\def\beq{\begin{equation}}   \def\eeq{\end{equation}}

\def\beeq{\begin{eqnarray}} \def\eeeq{\end{eqnarray}}

\begin{document}

\title{{\Large Fluctuations and Transverse Momentum Distributions in the Color Clustering Approach }}

\bigskip
\bigskip
\author{\large  L. Cunqueiro, J. Dias de Deus, E. G. Ferreiro, C. Pajares }
\email{elena@fpaxp1.usc.es}
\affiliation{Departamento de F\'{\i}sica de Part\'{\i}culas,
Universidade de Santiago de Compostela,
15782 Santiago de Compostela, Spain}
\bigskip
\bigskip

\begin{abstract}
\leftskip1.0cm
\rightskip1.0cm
We present our results on transverse momentum fluctuations,
multiplicity fluctuations
and transverse momentum distributions for baryons and mesons 
in the framework of the clustering of color sources.
We  determine under what conditions the
initial state configurations can lead to color connection, and more
specifically, if variations of the initial state can lead to a transition
from disconnected to connected color clusters, modifying
the number of effective sources.
We find that beyond a critical point, one has a condensate, containing
interacting and hence color-connected sources.
This point thus specifies the onset of color
deconfinement.
We show that 
the transverse momentum and multiplicity distributions are related
to
each other in a defined way.
We obtain a non-monotonic dependence of the $p_T$ and multiplicity fluctuations
with the number of participants.
We present our results for the fluctuations and the transverse momentum
distributions at RHIC energies compared to the existing experimental data
and our predictions for LHC energies.
\end{abstract}

\maketitle

\section{Introduction}

The main idea in the framework of the clustering of color sources is the fact 
that 
elementary color sources -strings- can overlap forming
clusters,
so the number of effective sources is modified. The procedure is the following:
in each collision,  color strings are
stretched between
the colliding partons. Those strings act as color sources of particles
which
are
successively
broken by creation of $q {\bar q}$ pairs from the sea.
The color strings correspond to small areas
 in the transverse space filled
with the color field created by the
colliding partons.
If the density of strings increases, they overlap in the transverse
space, giving rise to a phenomenon of string fusion and
percolation \cite{Armesto:1996kt}.
Percolation indicates that the cluster size diverges,
reaching the size of the system.

Thus, variations of the initial state can lead to a
transition from
disconnected to connected color clusters. The percolation point signals the
onset of color deconfinement.

These clusters decay into particles with mean transverse momentum and
mean multiplicity that
depend on the
number of elementary sources that conform each cluster, and the area
occupied by the cluster.
For a cluster of $n$ overlapping strings, the vectorial sum of the color charges of the individual strings leads to the following values for the total color charge, the mean multiplicity and the mean transverse momentum:
\begin{equation}
Q_n=\sqrt{\frac{nS_n}{S_1}}Q_1\ \ \ \
\mu_n=\sqrt{\frac{nS_n}{S_1}}\mu_1\ \ \ \
\langle p_T^2\rangle_n=\sqrt{\frac{nS_1}{S_n}}\langle p_T^2\rangle_1\, .
\end{equation}
Moreover, it is possible to obtain the 
following analytic expression that relates the areas with the density of 
strings $\eta$:
\beq
<\frac{n S_1}{S_n}>= \frac{\eta}{1-\exp{(-\eta)}} \equiv \frac{1}{F(\eta)^2}\, .
\eeq

\section{Transverse momentum distributions}

The clustering reduces the average multiplicity and
enhances the average $<p_{T}>$ of an event in a factor $F(\eta)$ with respect
to those resulting from pure superposition of strings :
  \begin{equation}<\mu>=N_{s}F(\eta) <\mu>_{1}, \;
  <p_{T}^{2}>=<p_{T}^{2}>_{1}/{F(\eta)} \label{average} \end{equation}
where $N_{S}$ is the number of strings and $F(\eta)=\sqrt{\frac{1-e^{-\eta}}{\eta}}$ is a function of the density
of strings $\eta$.

The invariant cross section can be written as a superposition of the
transverse momentum distributions of each cluster, $f(x,p_{T})$ --Schwinger
formula for the decay of a cluster--, weighted by
the distribution of the different tensions of the clusters, $W(x)$. This weight
function behaves as
a gamma function whose width is proportional to $1/k$ where $k$ is a
determined function of $\eta$ related to the measured dynamical transverse
momentum and multiplicity fluctuations \cite{DiasdeDeus:2003ei,Cunqueiro:2005hx}:
\begin{equation}\hspace{-1cm}\frac{dN}{dp_{T}^2 dy}=\int_{0}^{\infty}dx W(x)
f(p_{T},x)=\frac{dN}{dy}\frac{k-1}{k}\frac{1}{<p_{T}^2>_{1i}}F(\eta)\frac{1}{(1+\frac{F(\eta)p_{T}^{2}}{k<p_{T}^2>_{1i}})^{k}}.
 \label{spectra} \end{equation}

For (anti)baryons the equation (\ref{average}) must be changed to
$<\mu_{\overline{B}}>=N_{S}^{1+\alpha}F(\eta_{\overline{B}})<\mu_{1\overline{B}}>$
to take into account the fact that baryons are enhanced over mesons in the fragmentation of a high density cluster \cite{Cunqueiro:2007fn}. The parameter $\alpha=$0.09 is fixed from the
experimental dependence of $\frac{\overline{p}}{\pi}$ on $N_{part}$.
The (anti)baryons probe  higher densities than mesons,
$\eta_{B}=N_{S}^{\alpha}\eta$.  On the other hand, from the constituent
counting rules applied to the high $p_{T}$ behavior we deduce that for
baryons $k_{B}=k(\eta_{B})+1$.
In Fig. 1, we show the ratios $R_{CP}$ and $\frac{\overline{p}}{\pi^{0}}$
compared to RHIC experimental data for pions and antiprotons together with the LHC predictions.
In Fig. 2 (left) we show  the nuclear modification factor $R_{AA}$ for
pions  and protons  for central collisions at RHIC. LHC predictions are also shown.
We note that $p+p$ collisions at LHC energies will reach enough string density
for
nuclear like effects to occur. Because of this, in Fig. 2 (right) we show the ratio $R_{CP}$ for
$pp\to \pi X $ as a function of $p_{T}$, where the denominator is given by the
minimum bias inclusive cross section and the numerator is the inclusive cross
section corresponding to events with twice the multiplicity of 
the minimum bias one.
According to our formula (\ref{spectra}) a suppression at large $p_{T}$
occurs.

\begin{figure}
\begin{minipage}[t]{6cm}
\epsfig{figure=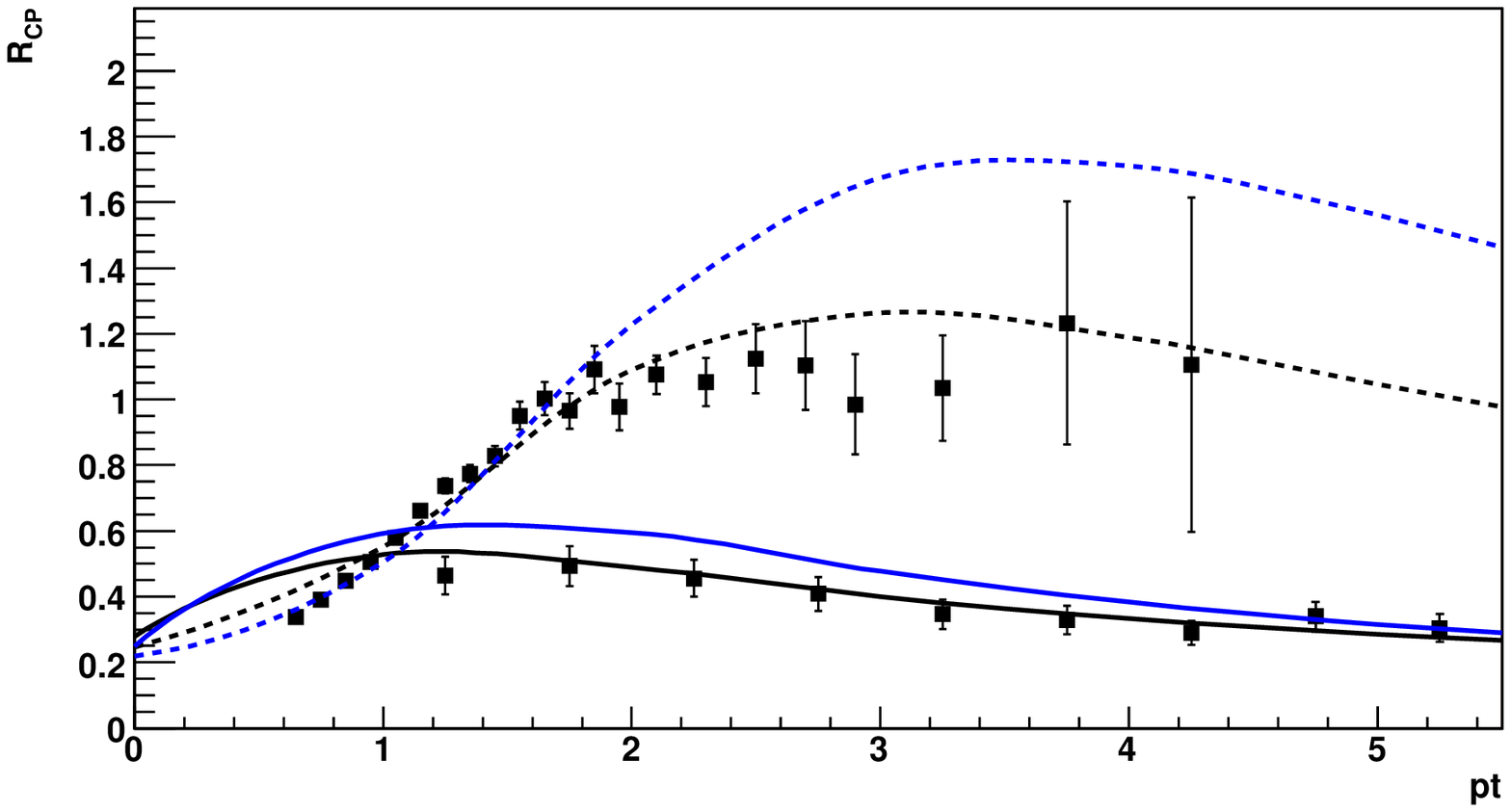,width=7cm,height=6cm}
\end{minipage}
\hfill
\begin{minipage}[t]{6cm}
\hspace{-2cm} \epsfig{figure=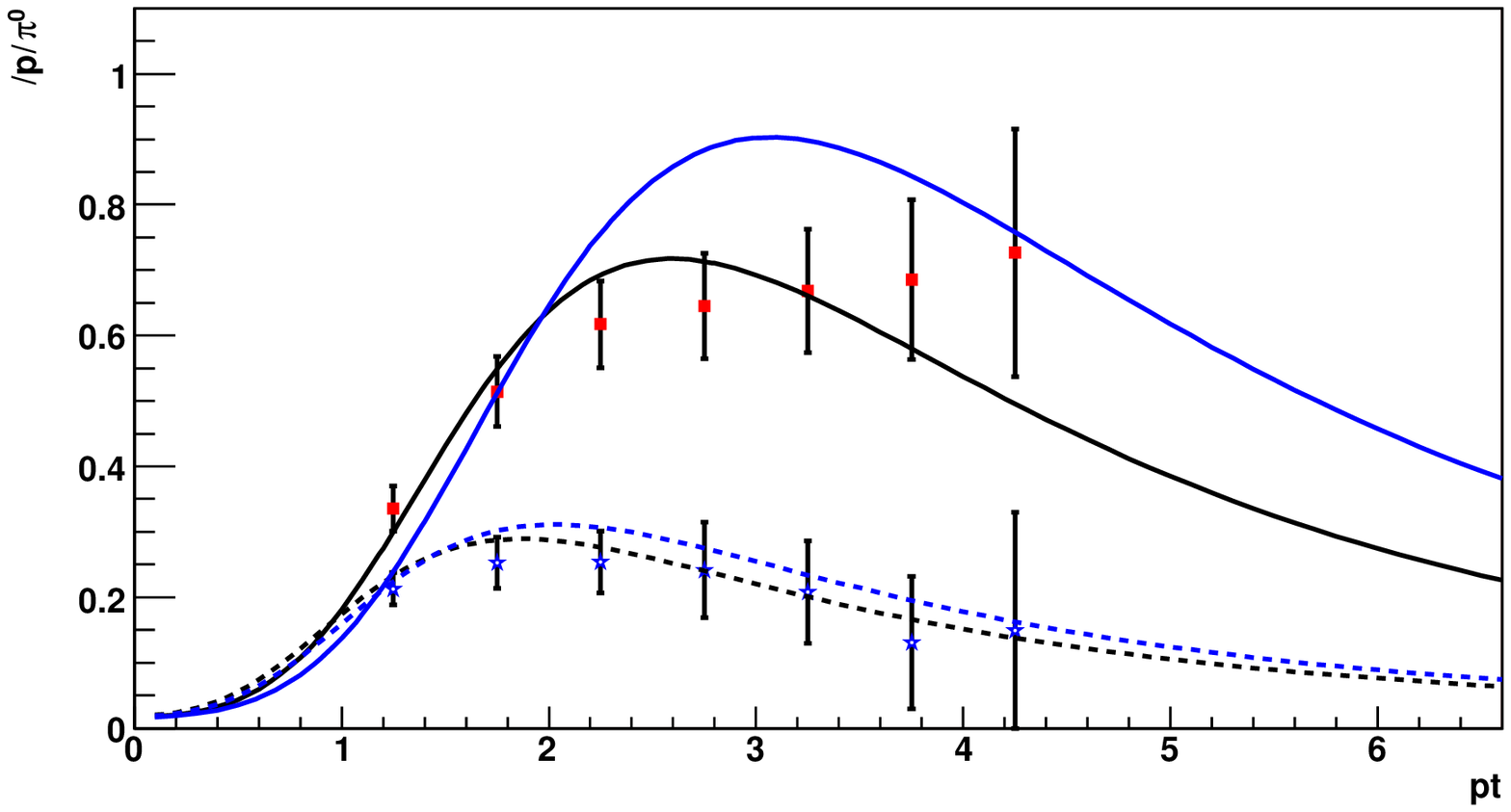,width=7cm,height=6cm}
\end{minipage}
\caption{Left: $R_{CP}$ for neutral pions (solid) and
  antiprotons (dashed). Right: $\overline{p}$ to $\pi^{0}$ ratio for the centrality
  bins 0-10\% (solid) and 60-92\% (dashed). RHIC results in black and LHC predictions in blue. }
\end{figure}
\begin{figure}
\begin{minipage}[t]{6cm}
\epsfig{figure=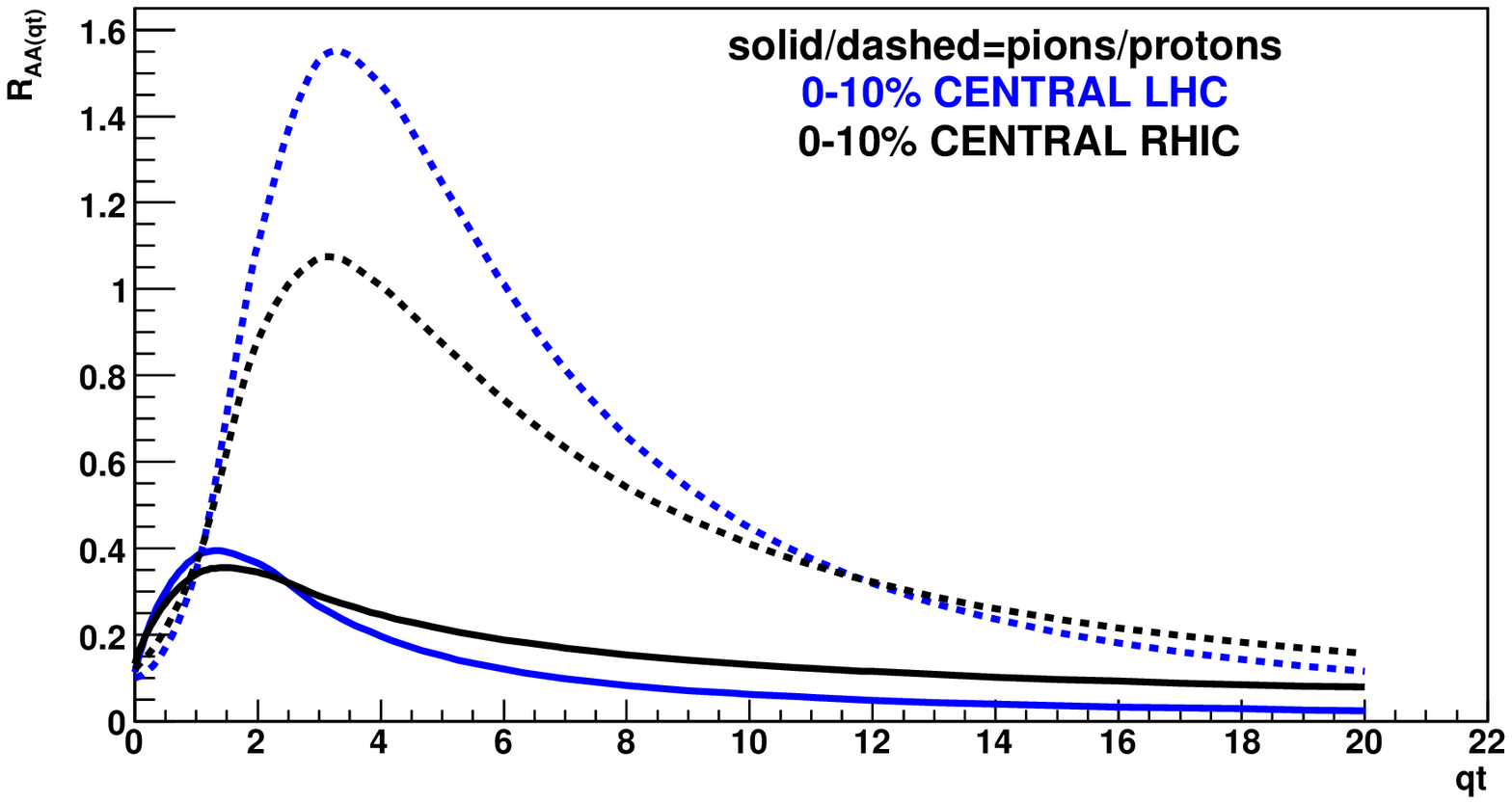,width=7cm,height=6cm}
\end{minipage}
\hfill
\begin{minipage}[t]{6cm}
\hspace{-2cm} \epsfig{figure=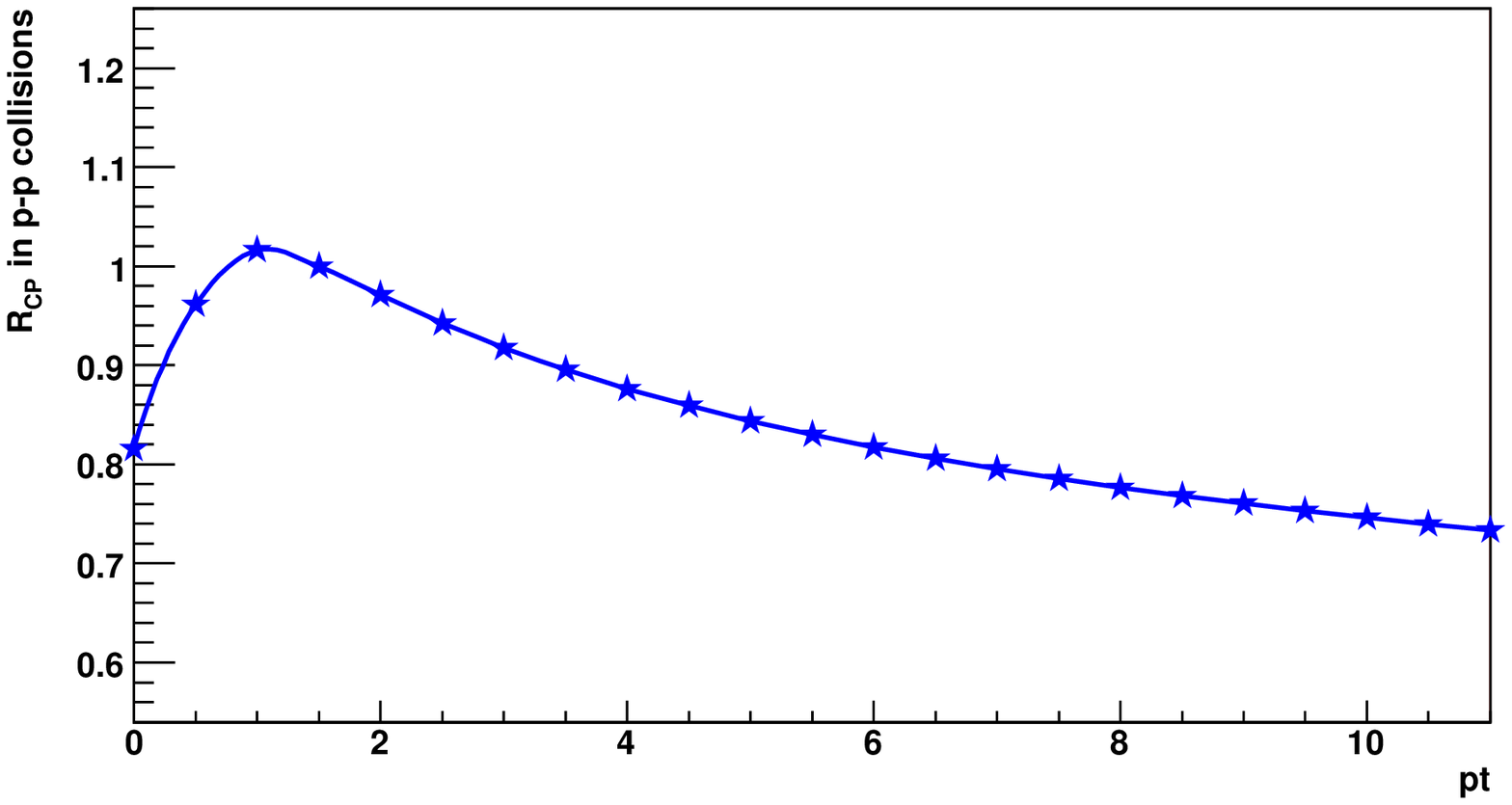,width=7cm,height=6cm}
\end{minipage}
\caption{Left: Nuclear Modification Factor for $\pi^{0}$ (solid) and
  $\overline{p}$ (dashed) for 0-10\% central events, RHIC results in
 black and LHC predictions in blue. Right: $R_{CP}$ for pions in p-p collisions at LHC. }
\end{figure}

\section{Fluctuations}

Non-statistical event-by-event fluctuations in relativistic
heavy ion collisions have been proposed as a probe of phase instabilities
near de QCD phase transition.
These
fluctuations show a non-monotonic behavior with the centrality of the
collision: they grow as the centrality increases, showing a maximum
at mid centralities, followed by a decrease at larger centralities.
Different mechanisms
have been proposed in order to explain those data. 
In the string clustering approach, 
the behavior of the $p_T$ \cite{Ferreiro:2003dw}
and multiplicity \cite{Cunqueiro:2005hx}
fluctuations can be
understood as
follows:
at low density, most of the particles are
produced by individual strings with
the same transverse momentum $<p_T>_1$ and the same multiplicity
$<\mu_1>$, so fluctuations are small.
At large density,
above the critical point of percolation, we have only one cluster, so
fluctuations are not expected either.
Just below the percolation critical density, we have
a large number of clusters formed by different number of strings $n$,
with
different size and thus different $<p_T>_n$
and different $<\mu>_n$
so the fluctuations are maximal.

The variables to measure event-by-event $p_T$ fluctuations are $\phi$ and
$F_{p_T}$,
that quantify the deviation of the observed fluctuations from
statistically independent particle emission:
\beq
\phi=\sqrt{\frac{<Z^2>}{<\mu>}}-\sqrt{<z^2>}\ ,
\eeq
where $z_i={p_T}_i - <p_T>$ is defined for each particle and
$Z_i=\sum_{j=1}^{N_i} z_j$ is defined for each event,
and
\beq
F_{p_T} = \frac{\omega_{data} - \omega_{random}}{\omega_{random}},\, \ \ \
\omega= \frac{\sqrt{<p_T^2>-<p_T>^2}}{<p_T>}\ .
\eeq
Moreover,
in order to measure the multiplicity fluctuations, the variance of the multiplicity distribution scaled to the mean value of
the multiplicity has been used.
Its behavior is similar to the one obtained
for $\Phi (p_T)$, used to quantify the $p_T$-fluctuations,
suggesting that they are related to each other.
The $\Phi$-measure is independent of the distribution of number
of particle sources
if the sources are identical and independent from each other. That is,
$\Phi$ should be
independent of the impact parameter if the nucleus-nucleus collision is
a simple superposition of nucleon-nucleon interactions.

\begin{figure}
\begin{minipage}[t]{6cm}
\epsfig{figure=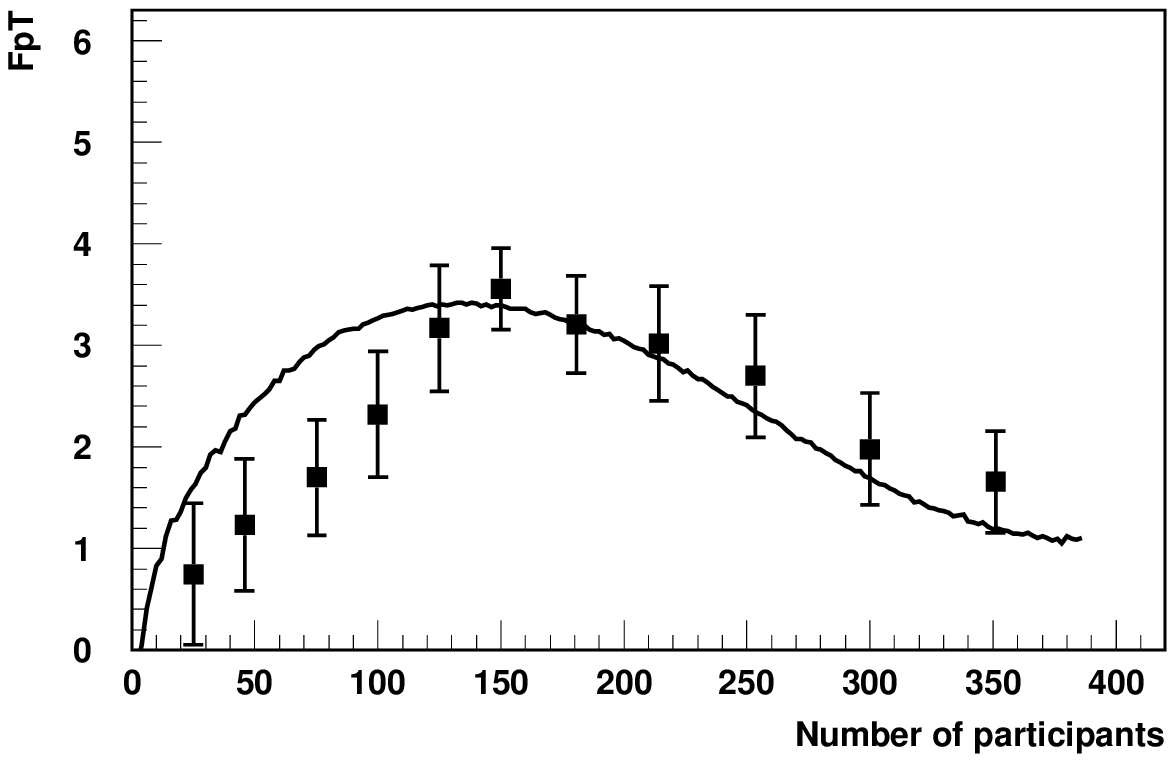,width=7cm,height=7cm}
\end{minipage}
\hspace{-3cm} 
\hfill
\begin{minipage}[t]{6cm}
\epsfig{figure=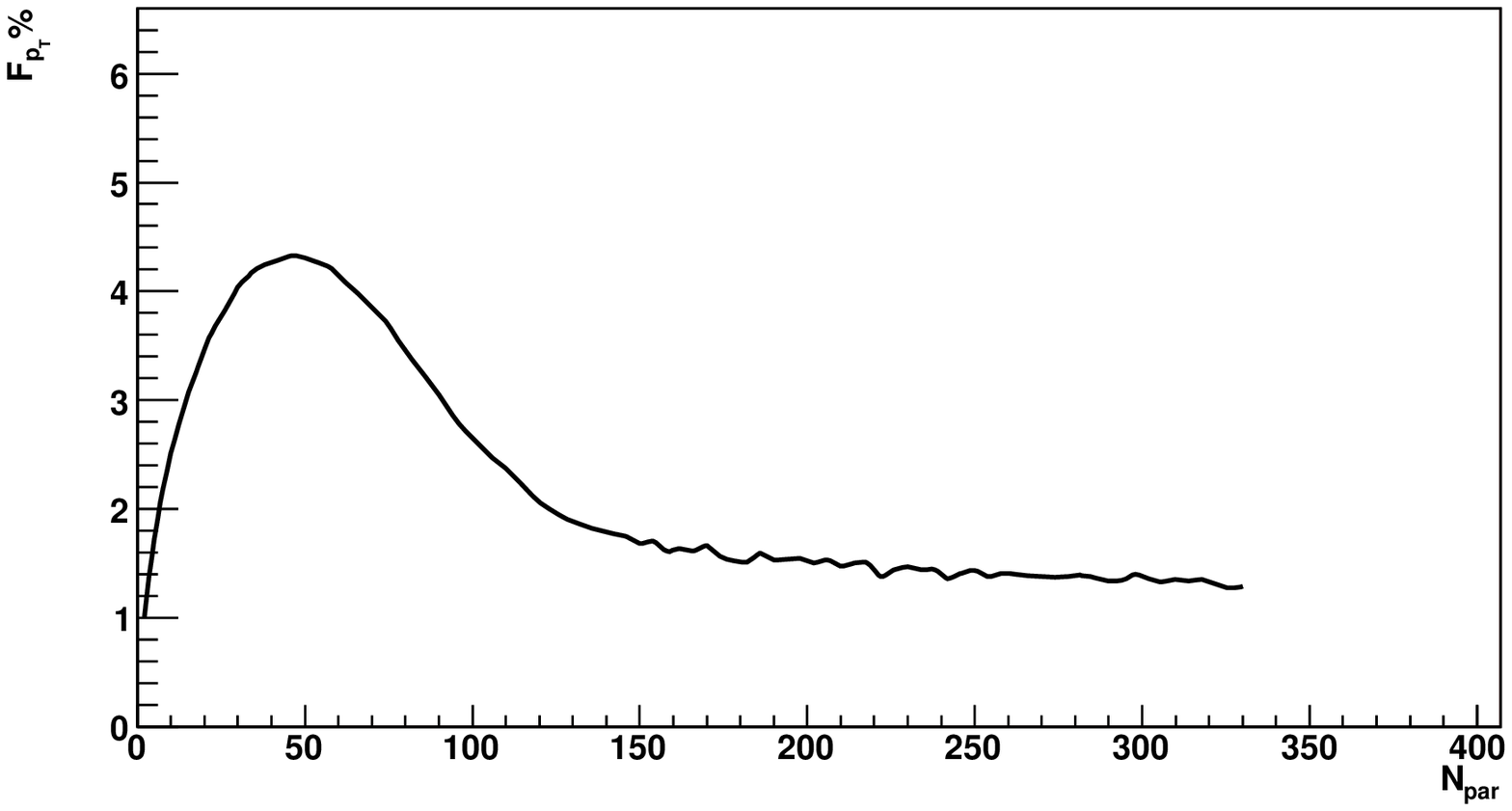,width=7cm,height=6cm}
\end{minipage}
\caption{
$F_{p_T} (\%)$
versus the number of participants at RHIC (left) and LHC (right) energies.
}
\end{figure}

\begin{center}
\begin{figure*}
\epsfxsize=8.5cm
\epsfysize=6.0cm
\centerline{\epsfbox{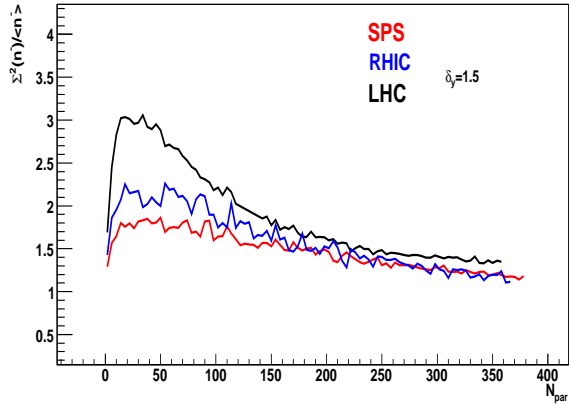}}
\vskip -0.25cm
\caption{Scaled variance on negatively charged particles at, from up to down,
LHC, RHIC and SPS.}
\vskip -0.5cm
\end{figure*}
\end{center}

In Fig. 3 we present our results on $p_T$ fluctuations at LHC.
Note that the increase of the energy essentially
shifts the maximum position to a lower number of participants
\cite{Ferreiro:2003dw}.
In Fig. 4 we show our values for the scaled variance of negatively charged particles
at SPS, RHIC and LHC energies.

\end{document}